\documentclass[aps,prb,twocolumn,longbibliography,groupedaddress]{revtex4-1}

\usepackage{hyperref}

\usepackage{graphicx}
\usepackage{dcolumn}
\usepackage{bm}
\usepackage{color}

\begin{document}



\title[Cubic anisotropy in high homogeneity thin (Ga,Mn)As layers]{Cubic anisotropy in high homogeneity thin (Ga,Mn)As layers}

\author{M. Sawicki}
 \affiliation{Institute of Physics, Polish Academy of Sciences,
Aleja Lotnikow 32/46, PL-02668 Warsaw, Poland}

\author{O. Proselkov}
 \affiliation{Institute of Physics, Polish Academy of Sciences,
Aleja Lotnikow 32/46, PL-02668 Warsaw, Poland}
\author{C. Sliwa}
 \affiliation{Institute of Physics, Polish Academy of Sciences,
Aleja Lotnikow 32/46, PL-02668 Warsaw, Poland}

\author{J. Sadowski}
\affiliation{Institute of Physics, Polish Academy of Sciences,
Aleja Lotnikow 32/46, PL-02668 Warsaw, Poland}
\affiliation{MAX-IV laboratory, Lund University, P.O. Box 118, SE-22100 Lund, Sweden}
\affiliation{Department of Physics and Electrical Engineering, Linnaeus University,   SE-391 82 Kalmar, Sweden}
\author{P. Aleshkevych}
 \affiliation{Institute of Physics, Polish Academy of Sciences,
Aleja Lotnikow 32/46, PL-02668 Warsaw, Poland}
\author{J.Z. Domagala}
 \affiliation{Institute of Physics, Polish Academy of Sciences,
Aleja Lotnikow 32/46, PL-02668 Warsaw, Poland}
\author{T. Dietl}
\affiliation{International Research Centre MagTop, Aleja Lotnik\'ow 32/46, PL-02668 Warsaw, Poland}
\affiliation{Institute of Physics, Polish Academy of Sciences,
Aleja Lotnikow 32/46, PL-02668 Warsaw, Poland}
\affiliation{WPI-Advanced Institute for Materials Research, Tohoku University, Sendai 980-8577, Japan}

\date{\today}

\begin{abstract}
Historically, comprehensive studies of dilute ferromagnetic semiconductors, e.g., $p$-type
(Cd,Mn)Te and (Ga,Mn)As, paved the way for a quantitative theoretical description of
effects associated with spin-orbit interactions in solids, such as crystalline
magnetic anisotropy. In particular, the theory was successful in explaining
{\em uniaxial} magnetic anisotropies associated with biaxial strain and non-random
formation of magnetic dimers in epitaxial (Ga,Mn)As layers. However, the situation appears
much less settled in the case of the {\em cubic} term: the
theory predicts switchings of the easy axis between in-plane $\langle 100\rangle$
and $\langle 110\rangle$ directions as a function of the hole
concentration, whereas only the $\langle 100\rangle$ orientation has been found
experimentally. Here, we report on the observation of such switchings by magnetization
and ferromagnetic resonance studies on a series of high-crystalline quality
(Ga,Mn)As films. We describe our findings by the mean-field $p$-$d$ Zener model
augmented with three new ingredients. The first one is a scattering broadening
of the hole density of states, which reduces significantly the amplitude
of the alternating carrier-induced contribution. This opens
the way for the two other ingredients, namely the so-far disregarded single-ion
magnetic anisotropy and disorder-driven non-uniformities of the carrier density,
both favoring the $\langle 100\rangle$ direction of the apparent easy axis.
However, according to our results, when the disorder gets reduced a switching to the $\langle 110\rangle$
orientation is possible in a certain temperature and hole concentration range.


\end{abstract}

\maketitle


\section{Introduction}

The discovery of carrier-mediated ferromagnetism in (III,Mn)V and (II,Mn)VI semiconductors systems makes it possible to examine the interplay between physical properties of semiconductor quantum structures and ferromagnetic materials \cite{Dietl:2014_RMP}.
At the same time, complementary resources of these systems allowed for novel functionalities and devices enabling magnetization manipulation \cite{Dietl:2014_RMP,Ohno:2000_N,Chiba:2008_N,Jungwirth:2014_RMP,Matsukura:2015_NN}, paving the way towards the industrial development stage for all-metal devices \cite{Matsukura:2015_NN,Alzate:2012_IEDM}.
In this context (Ga,Mn)As has served as a valuable test ground for new concepts and device architecture, due to the relatively high Curie temperature  $T_{\mathrm{C}}$  and its compatibility with the well-characterized GaAs system.
Importantly, despite much lower spin and carrier concentrations compared to ferromagnetic metals, (III,Mn)V dilute ferromagnetic semiconductors (DFS) exhibit excellent micromagnetic characteristics, including well defined magnetic anisotropy and large ferromagnetic domains.
The theoretical understanding of these materials is built on the $p$--$d$ Zener model of ferromagnetism \cite{Dietl:2000_S}.
In this model, the thermodynamic properties are determined by the valence band carriers
contribution to the free energy of the system, which is calculated taking the spin-orbit interaction into account within the $\bm{k} \cdot \bm{p}$ theory \cite{Dietl:2000_S,Dietl:2001_PRB,Jungwirth:2006_RMP,Zemen:2009_PRB,Stefanowicz:2010_PRBb}
or tight binding model \cite{Werpachowska:2010_PRBb,Werpachowska:2010_PRBa} with the $p$-$d$ exchange interaction between the carriers and the localized Mn spins considered within the virtual-crystal and molecular-field approximations.
In this approach the long-range ferromagnetic interactions between the localized spins are mediated by delocalized holes in the weakly perturbed $p$-like valence band \cite{Kepa:2003_PRL}.

The model explains well the influence of epitaxial strain on magnetic anisotropy and various experimentally observed magnetic easy axis reorientation transitions (SRT) as a function of temperature $T$ and hole concentration $p$ with a sound exception of the fourfold (cubic-like) component of the magnetic anisotropy for which neither a strong oscillatory dependence $\langle 100 \rangle \leftrightarrow \langle 110 \rangle$ on $p$ and $T$ (through magnetization, $M$) nor its predicted strength have been verified \cite{Sawicki:2003_JSNM,Sawicki:2004_PRB,Wang:2005_PRL,Zemen:2009_PRB,Glunk:2009_PRB,Stefanowicz:2010_PRBb}.
Intriguingly, only have the $\langle 100 \rangle$ in-plane cubic easy axis directions been reported in (Ga,Mn)As epilayers so far \cite{Tang:2003_PRL,Titova:2005_PRB,Hamaya:2006_PRBb,Liu:2006_JPCM,Pappert:2007_NJP,Thevenard:2007_PRB,Gould:2007_NJP,Glunk:2009_PRB}.



In this study we provide experimental evidences that the  $\langle110\rangle$ in-plane directions can become the cubic easy axes in (Ga,Mn)As.
These observations stem from the examination of magnetization curves and angular dependencies of ferromagnetic resonance (FMR) of carefully selected and prepared thin (Ga,Mn)As layers.
Interestingly, the $\langle 110 \rangle$ cubic easy axes are observed only in limited ranges of $p$ and $T$, indicating an oscillating nature ($\langle100\rangle \leftrightarrow \langle110\rangle$ switching) of the cubic anisotropy as a function of $T$ and/or on $p$.

The elaborated here effect is of a significant supportive value for the $p$--$d$ Zener approach to ferromagnetism of DFS in general and for the (III,Mn)V family in particular.
It confirms perhaps the last experimentally unproven qualitative prediction of the model: the oscillatory behavior of the cubic easy axis \cite{Dietl:2001_PRB}.
On the other hand, we show that our experimental findings are richer than the model can describe, even in the advanced form developed here to incorporate contributions from the single-ion anisotropy of $S = 5/2$ Mn spin and the disorder.
So, to reconcile the \textsl{experimental} findings with the model computations 
we include semi-quantitatively into our data analysis the well established, but somehow largely disregarded fact that the assumption of an excellent magnetic homogeneity of very thin (Ga,Mn)As layers is not valid due to two space-charge layers which are formed at the material interfaces \cite{Nishitani:2010_PRB,Sawicki:2010_NP,Proselkov:2012_APL,Chen:2015_PRL}.
These interfacial charges deplete considerably the two near-the-interface regions of (Ga,Mn)As, introducing a certain amount of electrical disorder into even the best optimized samples.
Then, on the account of the increasing magnitude of the fluctuations in the local hole density of states \cite{Dietl:2008_JPSJ,Richardella:2010_S}, the long range ferromagnetic (FM) coupling expected in an ideal high-$p$ and an edges-less material, acquires in these regions a mesoscopic character and a superparamagnetic-like (SP-L) properties are added to the expected "ideal" magnetic response of the bulk (Ga,Mn)As films \cite{Sawicki:2006_JMMM}.
Basing on some heuristic experimental considerations our study convincingly show that this is the presence of this SP-L response, which magnetic characteristics greatly resemble the $\langle100\rangle$-easy axis cubic anisotropy behavior, that is most likely responsible for an apparent rotation of the cubic anisotropy to $\langle100\rangle$ direction at low $T$ and/or low $p$ in our samples.
In this context our study appears o be a sizable step to bridge the gap which has separated the experiment and theory in this field.

\section{Samples and Experiment}

\begin{figure}
\includegraphics[width=8.5 cm]{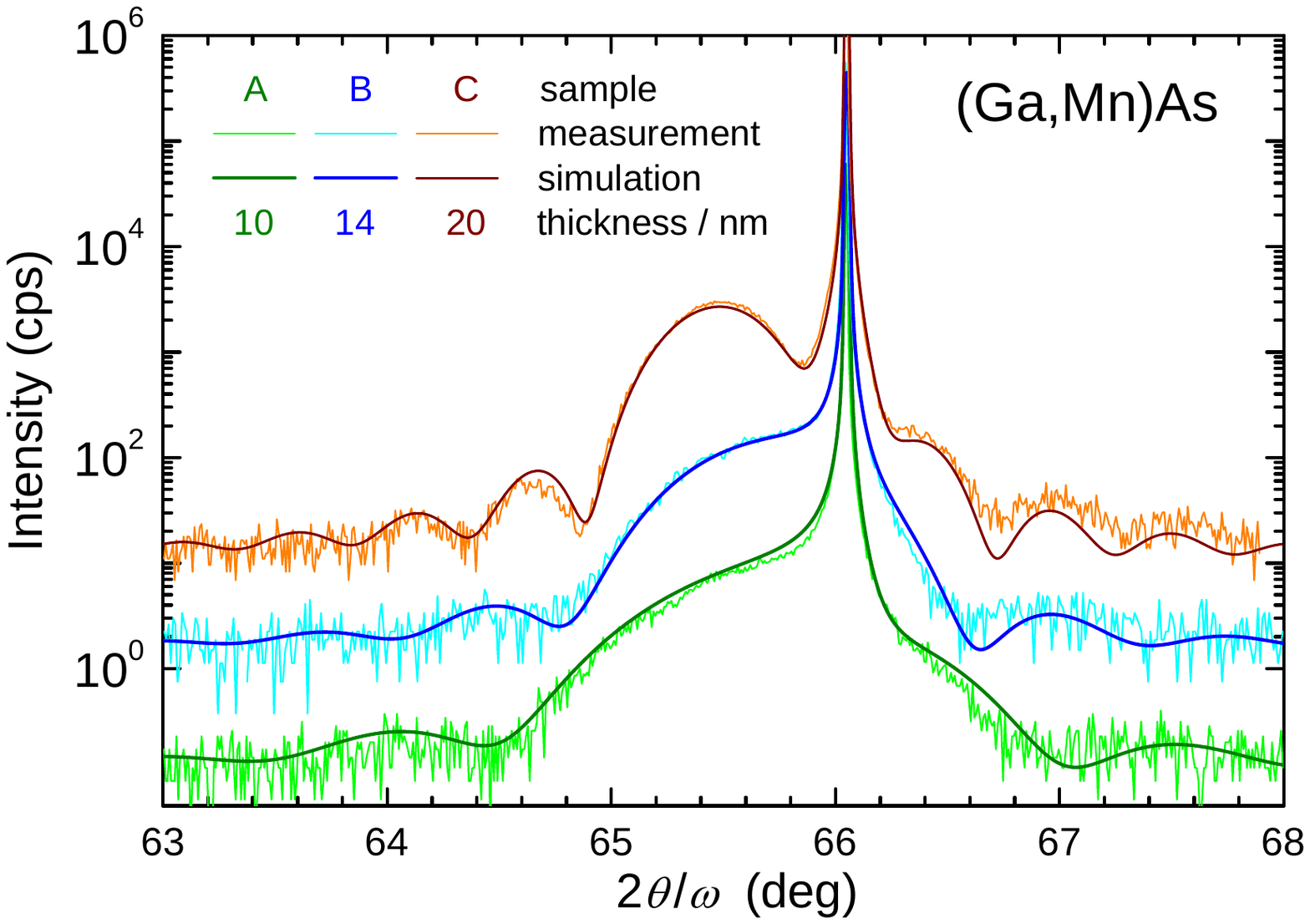}
 \caption{\label{Fig:XRD} (Color online) Fine lines of lighter shades mark high-resolution X-ray diffraction patterns of the studied (Ga,Mn)As layers: 004 Bragg reflections, $2\theta / \omega$ scans. 
 The central narrow features represent reflections from the GaAs substrate and the broader peaks at lower angles  are reflections from the layers. Thicker lines of darker shades mark simulations upon which Mn concentrations and the layers thicknesses have been established. }%
\end{figure}
A range of $10 < d < 20$~nm thin (Ga,Mn)As layers with Mn composition $x \simeq 10$\% has been deposited with a use of arsenic valve cracker effusion cell at 190~$^\circ$C by LT molecular beam epitaxy on about $18 \times 20$ mm$^2$ GaAs (100) substrates buffered with 20~nm thick LT-GaAs.
One of the layers have been subjected to \textit{in-situ} LT annealing under As capping \cite{Adell:2005_a}, whereas the rest of the layers are investigated in their as grown state or are subject to conventional open air oven LT annealing \cite{Edmonds:2004_PRL} at 180~$^\circ$C.
Their high structural quality has been confirmed by X-ray diffraction using laboratory Philips X-ray high resolution X'Pert MRD diffractometer with samples mounted on a high precision goniometric stage.
Figure~\ref{Fig:XRD} demonstrates $2\theta/\omega$ curves for the 004 Bragg reflections for  samples from this study.
As typically for (Ga,Mn)As deposited on GaAs substrates, the layers are fully strained, \textsl{i.e.} they have the same in plane lattice parameter as that of the substrate.
Diffraction peaks corresponding to the (Ga,Mn)As epitaxial layers shift to smaller angles with respect to that of the GaAs substrate, as a result of larger perpendicular lattice parameters.
Clear X-ray interference fringes  imply a high structural perfection of the layers and good quality of the interfaces.
Mn content and layer thickness $d$ are established upon simulation (marked as thicker solid lines of darker shades in Fig.~\ref{Fig:XRD}) performed using commercially available PANalytical EPITAXY software based on the dynamical theory of X-ray diffraction assuming elastic stiffness constants as for bulk GaAs and linear dependence of the lattice parameter of (Ga,Mn)As on $x$: $a(x)= 5.65469+0.24661x$ \cite{Sadowski:2001_APL}.
The results are listed in Table~\ref{tab:samples}.

Layers presented in this study have been selected according to their best lateral homogeneity, which has been assessed by $T_{\mathrm{C}}$ mapping across the substrate.
We note that whereas typical variations exceeding 5\% of $T_{\mathrm{C}}$  are observed in (Ga,Mn)As across similar substrates, in the three reported here layers, as indicated in Table~\ref{tab:samples}, the spread of their $T_{\mathrm{C}}$ values across the whole 2~cm substrate is smaller than 2~K, say 2\%.
Importantly, only in such layers cubic easy axes were found oriented along $\langle110\rangle$.

\begin{table}
\begin{tabular}{c c c c c}
\hline
Sample & Mn content & thickness & processing & $T_{\mathrm{C}}$\\
  &  \% &  nm &   & K\\
\hline
A  & 9 & 10 & \textsl{in situ} annealed & 153 \\ 
B  & 8.5 & 14 & as grown & 127 \\ 
C  & 10.5 & 20 & as grown & 95 \\ 
\hline
\end{tabular}
\caption{List of the (Ga,Mn)As layers investigated in this study for which the easy axes of the cubic magnetic anisotropy have been found to be aligned along $\langle 110 \rangle$ in-plane orientations. Mn content and thickness are determined upon X-ray diffraction pattern modeling. }
\label{tab:samples}
\end{table}


Magnetic measurements are carried out in a commercial MPMS XL Superconducting Quantum Interference Device (SQUID) magnetometer equipped with a low field option.
Customary cut long Si strips facilitate samples support in the magnetometer chamber \cite{Sawicki:2011_SST}.
A special demagnetization procedure has been employed to minimize the influence of parasitic fields during near-zero-fields measurements.
As the most relevant measurement are gathered at weak magnetic fields in ``hard axis" configuration we strictly follow the experimental code and data reduction detailed in Ref.~\onlinecite{Sawicki:2011_SST}. All the data presented here have their relevant diamagnetic contributions evaluated at room temperature and subtracted adequately.
The temperature dependence of remnant magnetization (TRM) measured along both cleaving edges of the sample ($\langle 110 \rangle$ directions for zinc blende substrates) serves to obtain an overview of magnetic anisotropy as well as to determine $T_{\mathrm{C}}$.
To study the magnetic anisotropy in a greater detail, magnetic hysteresis loops $M(H)$ are recorded in external magnetic field $H$ in the range of  $\pm 1$kOe along the same $\langle 110 \rangle$ in-plane directions.
The SQUID studies are supplemented by the in-plane angular dependence of the FMR performed at selected temperatures at  $\omega/2\pi = 9.3$~GHz.
It is shown below that both methods yield consistent results.

\section{Experimental Results}

\subsection{Overview of magnetic anisotropy}
\label{sec:MA}

There are few sources of magnetic anisotropy in (Ga,Mn)As epilayers: $T_d$ symmetry of the crystal, the epitaxial strain, a preferential aggregation of Mn dimers along one direction, and the shape anisotropy caused by the demagnetization effect.
Their sign and magnitude depend in turn on the ratio of valence band splitting to the Fermi energy, and so varies substantially with an effective Mn concentration ($x_{\text{eff}} = x_{\text{sub}} - x_{\text{I}} $, where $x_{\text{sub}}$ is the concentration of substitutional Mn at Ga sites and $ x_{\text{I}} $ is the concentration of interstitially located Mn species) and temperature (both determine the magnitude of $M$), epitaxial strain, and $p$.
It has been generally accepted that in layers where the easy plane configuration is realized, two leading terms are sufficient to adequately describe magnetization processes. 
They are the crystal symmetry related, fourfold (cubic) component \cite{Dietl:2001_PRB} and the Mn-dimers related uniaxial one \cite{Birowska:2012_PRL,Birowska:2017_PRB}.
So, the phenomenological description of the total in-plane magnetostatic energy of the system usually assumes the form:
\begin{equation}\label{Eq:energy}
E_{\text{m}} =\frac{K_C}{4}\sin^22\phi + K_U\sin^2\phi - M_SH \cos(\phi_H - \phi).
\end{equation}
Here,  $K_C$ and $K_U$ denote the lowest order cubic and uniaxial anisotropy constants,  $M_S$ is the saturation magnetization, and $\phi_H$ and $\phi$  are the angles of $H$ and $M$ to the $[100]$ direction. 
In its chosen form equation Eq.~(\ref{Eq:energy}) takes into account that both components are angled at $\pi/4$ with each other and that the positive sign of  $K_C$ represents $\langle 100 \rangle $ orientation of the easy exes of the cubic component.
It also gives an account for easy $\leftrightarrow$ hard axis switching represented here by a change of the sign of the relevant $K$.
These are the $[1\bar{1}0] \leftrightarrow [110]$ $\pi/2$ rotations for the uniaxial term \cite{Sawicki:2005_PRB,Chiba:2008_N}, and, for the very first time reported here, $\pi/4$ in-plane rotations, $\langle 100 \rangle \leftrightarrow \langle 110 \rangle$, of the cubic term.
Additionally, since magnitudes of $K_C$ and $K_U$ are $x_{\text{eff}}$, $p$, and most importantly, $T$-dependent, so yet another in-plane SRT frequently takes place at temperature where $K_U = K_C$ \cite{Wang:2005_PRL}.
This 2nd order magnetic SRT separates two different regimes.
When $|K_C| < |K_U|$, as it is the case of the present study, (Ga,Mn)As acquires nearly perfect magnetic uniaxial properties.
In such a case the presence of a weaker cubic term modifies the uniaxial behavior only very little.
In particular, the cubic does not reveal its presence at $ H=0$, and so to reveal its properties an external magnetic field is required.

The relationship between  the uniaxial hard axis magnetization $M_{[110]}$ and $H$ is obtained by minimizing the energy given by Eq.~(\ref{Eq:energy}) with respect to $\phi$ while setting $\phi_H = \pi/4$:
\begin{equation}\label{Eq:Hm110}
 H M = 2(K_U-K_C)\bar{m}_{[110]} + 4K_C \bar{m}_{[110]}^3,
 \end{equation}
where the reduced hard axis magnetization $\bar{m}_{[110]} = M_{[110]} / M$.
The first term in Eq.~(\ref{Eq:Hm110}), dominating  when $ \bar{m}_{[110]} \simeq 0$, that is at very weak magnetic fields, describes the initial, linear in $H$, magnetization process which begins with a slope $s= M/( K_U - K_C)/2$.
Here the influence of the fourfold anisotropy is only quantitative.
It redefines only the magnitude of the initial slope of the otherwise linear response.
However, it plays the decisive role at mid-field region (\textit{i.e.} when $\bar{m}_{[110]} \rightarrow 1$) since it sets both the strength and the curvature of the non-linear  part of $ \bar{m}_{[110]} (H)$.
Accordingly, $ \bar{m}_{[110]} (H)$ bends downwards for $K_C > 0$, exhibiting the typical concave character reported so far, but in the mid-field region it will turn upwards exhibiting the \textsl{convex} curvature for $K_C < 0$.
Actually, the initial reduction of $s$ by negative $K_C$ makes the convex curvature of $\bar{m}_{[110]}(H)$ a bit more pronounced, creating together an unmistakable fingerprint that the cubic easy axes assume $\langle 110 \rangle$ directions in the sample.

\subsection{Experimental determination of anisotropy constants}
\label{sec:ExpDet}

\begin{figure}
\includegraphics[width=8.5 cm]{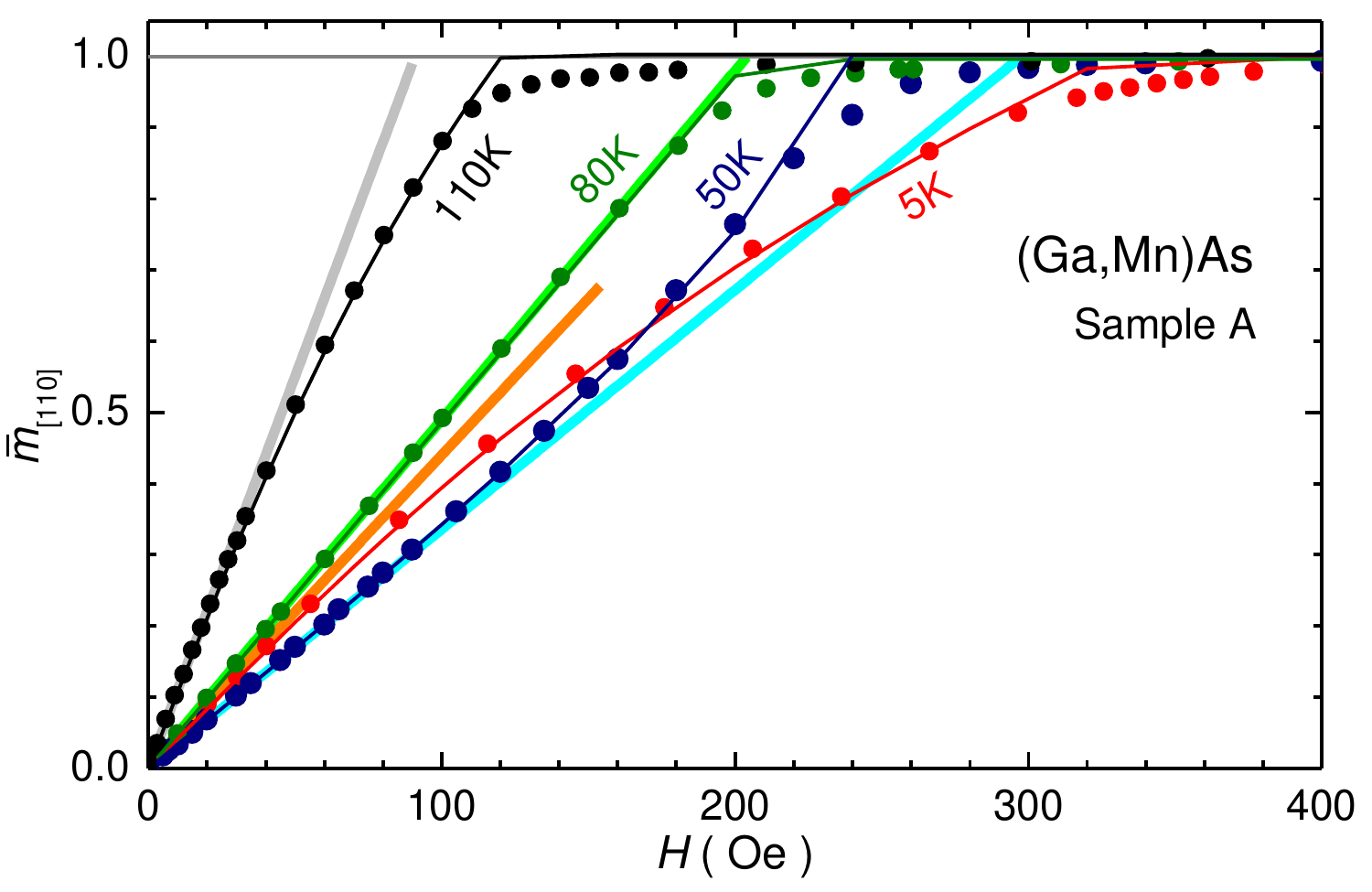}
 \caption{\label{Fig:ExampleM-H} (Color online) Uniaxial hard axis magnetic isotherms $\bar{m}_{[110]}(H) = M_{[110]}(H) / M$  at selected temperatures for sample A 
 (solid points) exemplifying different curvatures of mid--field part of the $m(H)$. The background thick lines of lighter shades indicate the initial slope of each $\bar{m}_{[110]}(H)$  and serve as references to ease the identification of the curvatures. The thin solid lines of matching colors are calculated from Eq.~(\ref{Eq:Hm110}) treating the anisotropy constants as adjustable parameters. }%
\end{figure}

The case of the convex curvature in $\bar{m}_{[110]}(H)$ is exemplified in Fig.~\ref{Fig:ExampleM-H} for sample A. 
Clearly, after a linear start, an up-turn is seen for $\bar{m}_{[110]}(H)$ taken at 50~K.
Interestingly, at 80~K the $ \bar{m}_{[110]}(H)$ remains linear in $H$ nearly up to the full saturation, indicating that around this temperature the curvature changes its sign, heralding the  SRT of the cubic anisotropy at around this $T$.
This fact is corroborated by a concave shape of $\bar{m}_{[110]}(H,110~K)$.
The existence of these different curvatures is highlighted in Fig.~\ref{Fig:ExampleM-H} by shaded thick background lines marking the initial slope of $\bar{m}_{[110]}(H)$ at these temperatures.

Interestingly, a similar to $T=$~110~K concave character of $\bar{m}_{[110]}(H)$ is seen at the lowest temperatures, exemplified in Fig.~\ref{Fig:ExampleM-H} for $T=$~5~K.
As argued above, to account for such $\bar{m}_{[110]}(H)$ a \textsl{positive} $K_C$ is required, but it will be reasoned further in the text that the incorporation of the previously disregarded in the analysis of magnetization curves in (Ga,Mn)As contribution from the SP-L component may well lead to the same concave curvature.

\begin{figure}
\includegraphics[width=\columnwidth]{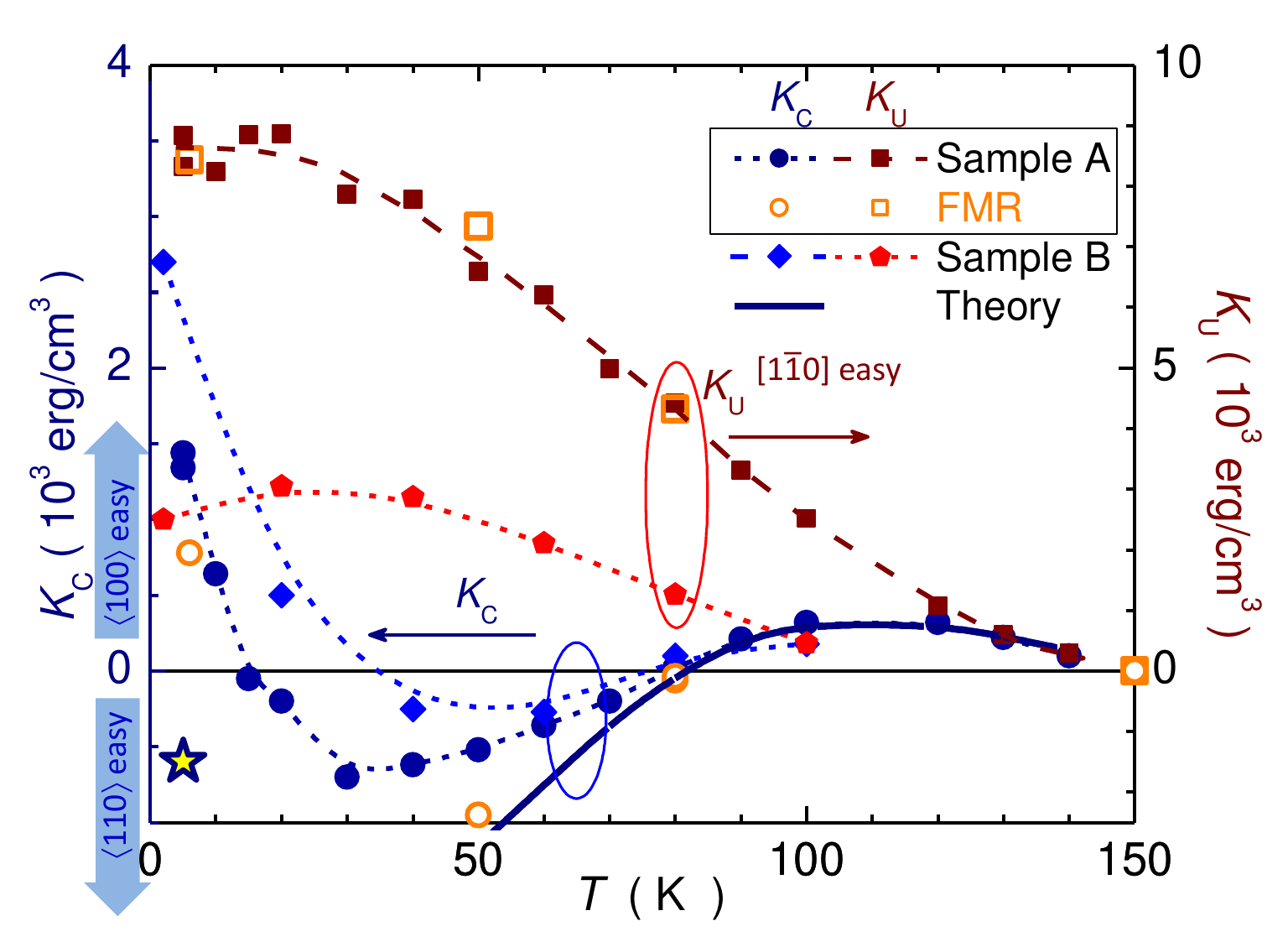}
 \caption{\label{Fig:KC-KU-T}  (Color online) Temperature dependence of uniaxial ($K_U$, squares and pentagons) and cubic ($K_C$, bullets and diamond) anisotropy constants. Solid points are obtained from analysis of uniaxial hard axis magnetization curves $ \bar{m}_{[110]}(H)$ open ones are obtained from analysis of angular dependence of FRM resonance positions. Dashed lines are guides for the eye only. In the notion adopted here the positive/negative sign of $K_C$ indicates that the cubic easy exes are aligned along $\langle 100 \rangle / \langle 110 \rangle$ directions, respectively. The star represents low temperature estimation of $K_C$ after removal from the original $ \bar{m}_{[110]}(H)$ a part attributed to nonhomogeneous magnetization originated at the interface regions, as detailed in section~\ref{sec:Inerface}.
 Thick solid line represent results of the Zener mean-field model, including a Gaussian broadening of the density of states and a single-ion  anisotropy contribution, as described in section~\ref{sec:TheoM}.}%
\end{figure}

A complete set of $ M_{[110]} (H)$ curves obtained in a broad temperature range permits us to establish upon Eq.~(\ref{Eq:Hm110}) the temperature dependence of the anisotropy constants in our samples.
To this end we take advantage that at each temperature both $K_{C}$ and $K_{U}$ are bound by the experimentally established magnitudes of $M$ and the initial slope $s$.
This constrain assures a perfect fit at weak field region 
and reduces the whole analysis to a simple choice of $K_{C}$ to reproduce the mid-field curvature of $ \bar{m}_{[110]}(H)$ for the already fixed $s$.
Such determined magnitudes of $K_{C}$ and $K_{U}$ are collected in Fig.~\ref{Fig:KC-KU-T}.
Interestingly, while exhibiting much lower amplitudes than $K_{U}$, the sign of cubic component clearly oscillates as a function of $T$.
On lowering $T$ the first rotation $\langle100\rangle \rightarrow \langle110\rangle$ takes place at around 80~K, and a rotation back to $\langle100\rangle$ at much lower temperatures is suggested by the analysis.


\subsection{Ferromagnetic resonance}

\begin{figure}
\includegraphics[width=8.5 cm]{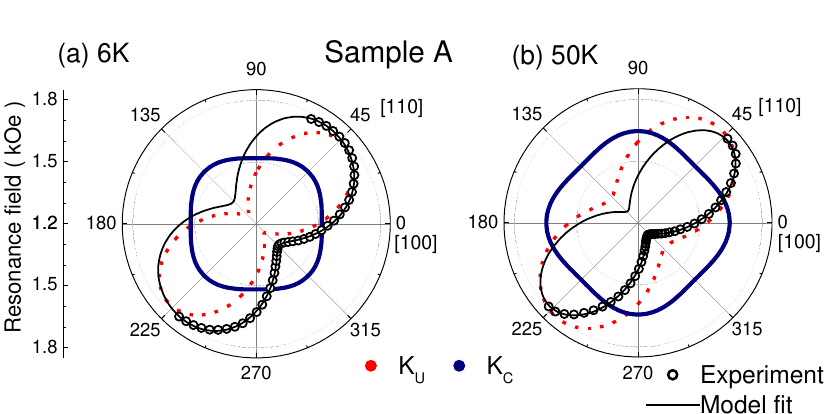}
 \caption{\label{Fig:FMR-polar}  (Color online) Polar plot of the FMR resonance position (open points) and the established uniaxial (doted red thick line) and cubic (navy thick line) contribution to the total magnetic energy at: 6 K (a)  and 50 K (b). }%
\end{figure}

This $\pi/4$ rotation of the cubic component on temperature is fully confirmed by the FMR studies.
In Fig.~\ref{Fig:FMR-polar} the dependence of the measured resonant fields on the orientation of the applied magnetic field is shown for the same sample A at 6 and 50~K (open circles).
The resonant field is obtained by evaluating the standard Artman equation \cite{Artman:1957_PR} at the equilibrium position of $M$ ($\partial E_ {\text{m}}/\partial\phi = 0$) treating $K_C$ and $K_U$ as fitting parameters.
The thin solid lines in Fig.~\ref{Fig:FMR-polar} show the established dependency of the resonant fields on $\phi_H$ along with  its decomposition into  uniaxial (doted red thick line) and cubic (dark thick line) contributions.
The extracted magnitudes of $K_C$ and $K_U$ are presented in Fig.~\ref{Fig:KC-KU-T} (open symbols) exhibiting a perfect correspondence to the results obtained from magnetization studies.
Again, analyzing the data within a frame set by Eq.~(\ref{Eq:energy}) we obtain a suggestive picture that the cubic anisotropy changes sign at the lowest temperatures.

\subsection{Dependence on hole density}

\begin{figure}
\includegraphics[width=\columnwidth]{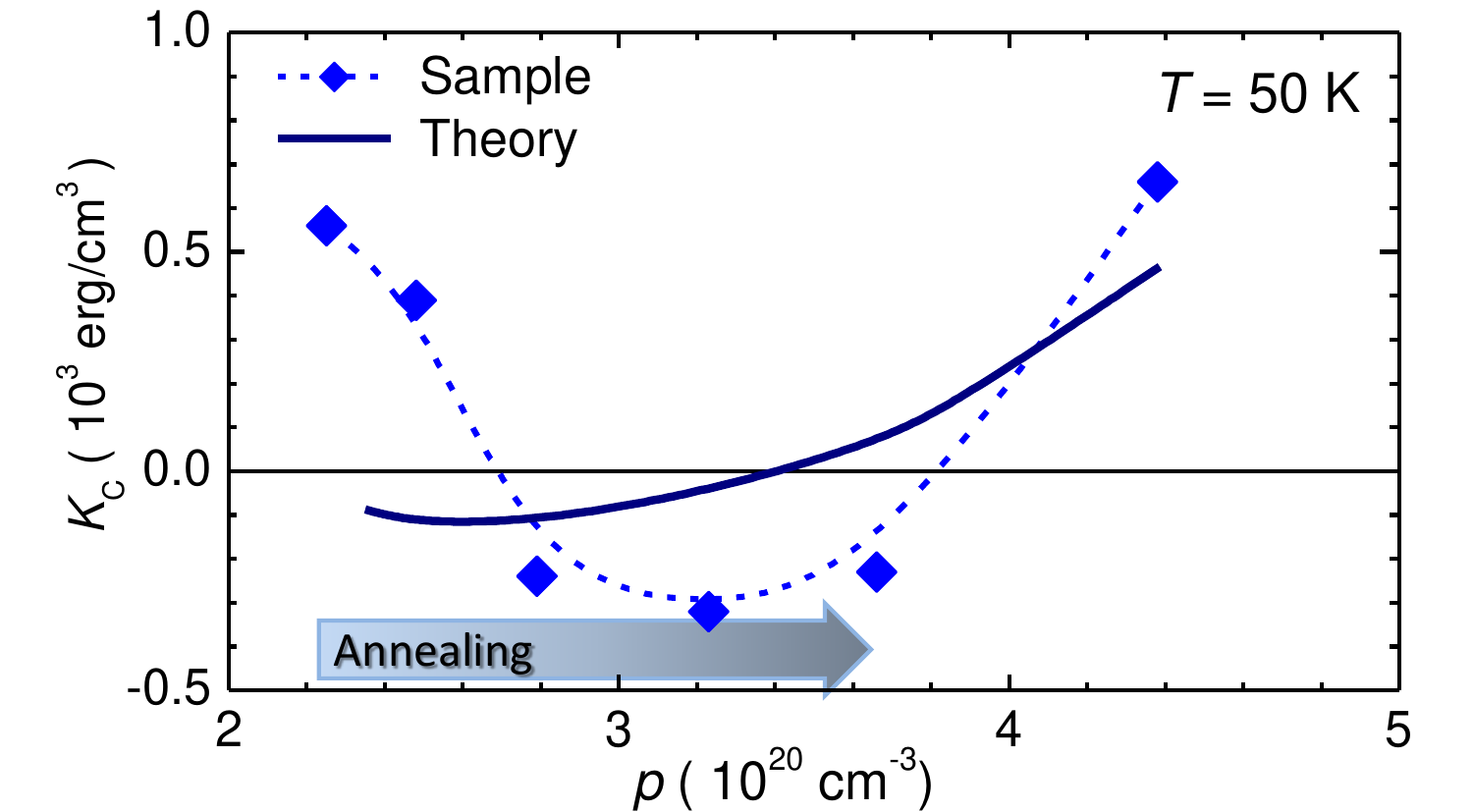}
 \caption{\label{Fig:KC-KU-(p)} (Color online) Hole density $p$ dependence of cubic  anisotropy constant ($K_C$, diamonds). Dashed line is a guide for the eye only. Thick solid line represents results of the Zener mean-field model including a Gaussian broadening of the density of states and a single-ion  anisotropy contribution, as detailed in section~\ref{sec:TheoM}.
}%
\end{figure}
Having established that the cubic easy exes can assume $\langle110\rangle$ in-plane directions in (Ga,Mn)As we can now turn to a more fundamental question of the role of hole density.
To this end we subject sample C to incremental open air LT annealing \cite{Hayashi:2001_APL,Edmonds:2002_APL}, since the corresponding out-diffusion and surface passivation
of Mn interstitials (Mn$_{\text{I}}$) increases  $p$ \cite{Wang:2004_JAP}.
We perform the annealing in small steps at progressively increasing temperatures (from 150 to 180 C) and annealing times: from 1  to 36 h.
To assess the changes in electrical and micromagnetic properties caused by the annealing, the full suite of magnetic measurements is performed after each annealing step.
Knowing the sample's $T_{\mathrm{C}}$ and saturation magnetization $M_{\mbox{\tiny S}}= x_{\text{eff}} N_0 g \mu_B S$ at each annealing step, the corresponding hole density $p = N_0(3x_{\text{eff}} - x)/2$, \textsl{i.e.} is neglecting other charge compensating defects, is computed in the framework of the mean-field $p-d$ Zener model, 
treating the problem in a self-consistent way by the incorporation of hole contribution to $M$ calculated for the same $x_{\text{eff}}$ \cite{Dietl:2000_S,Dietl:2001_PRB}.
Here $N_0 = 2.21 \times 10^{22}$~cm$^{-3}$ is the cation concentration in GaAs, $S = 5/2$ is the Mn spin, $g = 2.0$, and $\mu_B$ is the Bohr magneton.
We confirm that with fixed $x_{\text{sub}} = 9.3$\% this procedure allows us to reproduce exactly the experimentally established  magnitudes of $T_{\mathrm{C}}$ within 10\% margin for $M_{\mbox{\tiny S}}$.
In Fig.~\ref{Fig:KC-KU-(p)} the established at 50~K magnitudes of anisotropy constants are plotted as a function $p$, clearly indicating an oscillatory dependence of $K_C$ on $p$.
Remarkably, this is qualitatively the dependency that is predicted by the mean-field $p-d$ Zener model \cite{Dietl:2001_PRB},
however, contradictory to the model calculations (c.f. Fig. 9 in Ref.~\onlinecite{Dietl:2001_PRB}), the negative dip in $K_C(p)$  is much shallower and spans a narrower band in $p$. 

\section{Theoretical modeling}
\label{sec:TheoM}
Undoubtedly, the presented here overall qualitative agreement between the experimental findings and the theory of DFS based on the $p$--$d$ Zener model constitutes a great leap toward a reconciliation of experiment and model predictions.
On the other hand, the noted quantitative discrepancies call for a more in-depth reexamination of both the experimental and theoretical approaches.

We start our attempt from the theoretical side and introduce two ingredients to the standard theory of magnetic anisotropy in DFSs \cite{Dietl:2014_RMP,Jungwirth:2014_RMP}. First, we consider how scattering-induced broadening of density of states affects the amplitude of cubic magnetic anisotropy as a function of the hole concentration. Second, we examine the role of single-ion magnetic anisotropy. Our results demonstrate that the disorder-induced reduction in the magnitude of carrier-mediated magnetic anisotropy opens the floor for single-ion anisotropy despite its relatively small magnitude for Mn ions in the orbital singlet state. These two effects work together and elucidate why the $\langle 100 \rangle$ orientation of the easy axis is more frequently observed experimentally.

\subsection{Disorder effects}
\begin{figure}
	\includegraphics[width=0.95\columnwidth]{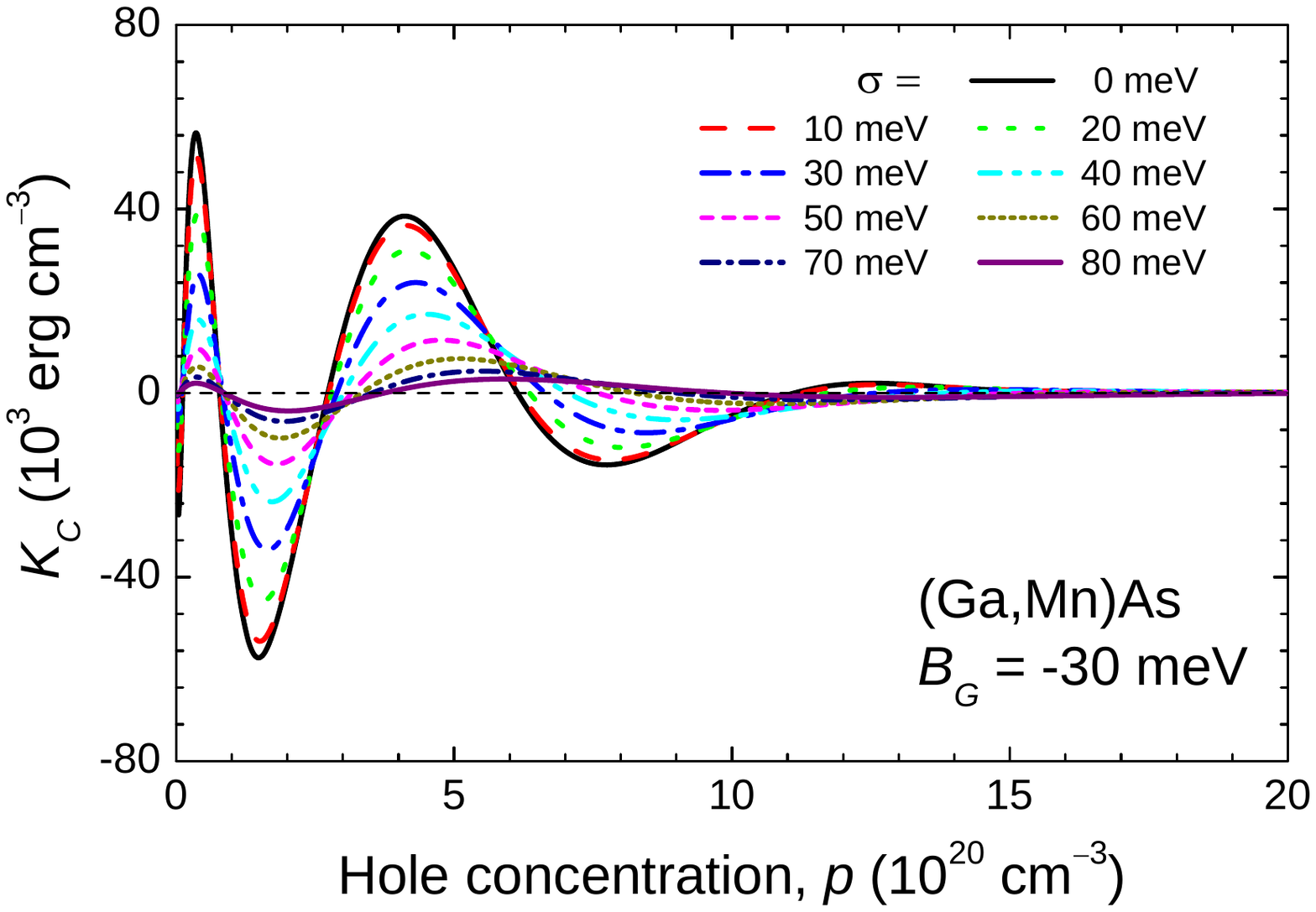}
	\caption{\label{fig: gaussian_broadening} (Color online) %
		The effect of disorder on the magnitude of the carrier-mediated in-plane cubic anisotropy coefficient $K_C$ computed  within
		 the mean-field $p$--$d$ Zener model extended by including a Gaussian broadening of the hole energy states
		for various values of the standard deviation~$\sigma$.}%
\end{figure}

In order to describe effects of disorder associated, in particular,  with the presence of randomly distributed ionized Mn
acceptors, we incorporate into the $p$--$d$ Zener model of magnetic anisotropy a Gaussian broadening of hole energy states,
the procedure employed previously in studies of Curie temperatures
in $p$-(Cd,Mn)Te quantum wells \cite{Kossacki:2000_PE}. As shown in Fig.~\ref{fig: gaussian_broadening}, such
 an approach predicts a reduction in the amplitude of the cubic anisotropy energy. This reduction
 is already twofold for the standard deviation $\sigma = 40$\,meV, i.e., for the lower bound value expected for the life
time energy broadening, typically, comparable to the Fermi energy in (Ga,Mn)As \cite{Dietl:2014_RMP}.

A decrease in the magnitude of the carrier-mediated term  enhances the relative importance of single-ion magnetic anisotropy.


\subsection{Single-ion magnetic anisotropy}

The cubic anisotropy of a single Mn spin with $S = 5/2$ is described by the Hamiltonian,
\begin{equation}
  \mathcal{H}_{SI} = \frac{a}{6} \left[ S_x^4 + S_y^4 + S_z^4
    - \frac{S(S+1)(3S^2+3S-1)}{5} \right].
\label{eq:SI_hamiltonian}
\end{equation}
where, according to electron paramagnetic resonance studies, $a \simeq -2.85 \times 10^{-19}$~erg in GaAs:Mn \cite{Bihler:2008_PRB,Fedorych:2002_PRB}.
The negative sign of $a$ implies the orientation of the cubic easy exes along
the $\left\langle 100 \right\rangle$ family of crystallographic directions.


In order to determine the magnitude of single-ion magnetic anisotropy in the low-temperature limit, we
calculate the expectation values $E^{SI}_{S}(\theta, \phi)$ of $\mathcal{H}_{SI}$
in the spin coherent states $\left|\psi^{SC}_{S}(\theta, \phi) \right\rangle$,
\begin{equation}
  E^{SI}_{5/2}(\theta, \phi) = \frac{4a}{125} \left[
    \langle S_x \rangle^4 + \langle S_y \rangle^4 + \langle S_z \rangle^4
    - \frac{375}{16} \right].
\label{eq:SI_LT}
\end{equation}
This expression represents the lowest order cubic anisotropy energy and for non-interacting Mn spins leads to the in-plane
anisotropy coefficient,
\begin{equation}
  K^{SI}_C = x_{\text{eff}} \, 15.6 \times 10^3 \, \mathrm{erg} / \mathrm{cm}^3.
\end{equation}

As seen by comparing Eqs.\,\ref{eq:SI_hamiltonian} and \ref{eq:SI_LT},  the low temperature quantum limit of $K^{SI}_C$ is about 5 times reduced with respect to the value expected for the classical vector $S =5/2$.

\begin{figure}
	\includegraphics[width=0.85\columnwidth]{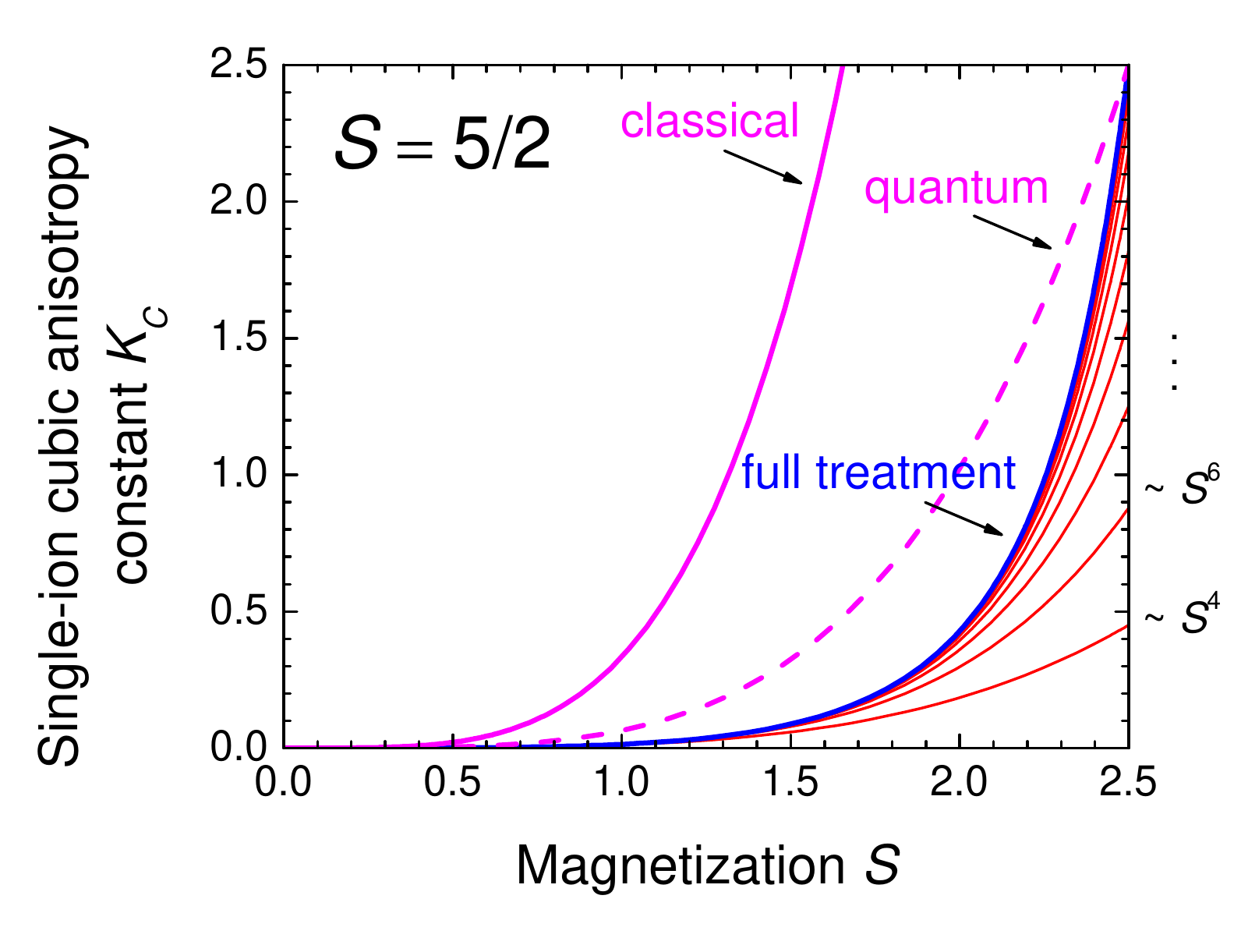}
	\caption{\label{fig: single_ion} (Color online) %
		Single-ion cubic anisotropy coefficient for non-interacting $S = 5/2$ spins as a function of magnetization
		$\left\langle S \right\rangle$, in units of $-a$. The curve designated ``classical''
		represents evaluation treating $S$ as a classical vector, while the dashed curve ``quantum'' is rescaled
		by the factor of $24/125$ to match the low-temperature limit of the quantum calculation. ``Full treatment''
		corresponds to the mean-field approach to the ferromagnetic case [Eq.\,\ref{eq:SI_MFA})].
        The thin red curves represent consequent orders of expansion
		in powers of magnetization, while the blue thick curve has been obtained by the Aitken extrapolation.}%
\end{figure}

We are interested in the role played by single-ion magnetic anisotropy in the case of ferromagnetic (Ga,Mn)As, i.e.,
in the presence of the hole liquid.
In order to evaluate the magnitude of $K^{SI}_C$ in such a case we take the magnetization
vector $\mathbf{M}$ of Mn spins as an order parameter and consider the
Landau-Ginzburg free energy
functional in the form containing the Mn contribution in the
absence of carriers and the carrier term \cite{Dietl:2014_RMP},
\begin{equation}
{\cal{F}}(\mathbf{M}) = {\cal{F}}_S(\mathbf{M}) + {\cal{F}}_{\text{c}}(\mathbf{M}),
\end{equation}
where
\begin{equation}
{\cal{F}}_{\text{S}}(\mathbf{M}) = \int_0^{\mathbf{M}} {\text{d}} \mathbf{M}_o\cdot\mathbf{h}(\mathbf{M}_o) - \mathbf{M}\cdot\mathbf{H}.
\label{eq:SI_MFA}
\end{equation}
Here, $\mathbf{h}(\mathbf{M}_o)$ denotes the inverse function to $\mathbf{M}_o(\mathbf{h})$, where $\mathbf{M}_o$ is the magnetization of Mn spins in the absence of carriers in the field $\vec{h}$ and temperature $T$ computed from the single-ion spin hamiltonian given in Eq.\,(\ref{eq:SI_hamiltonian}),
supplemented by the Zeeman term $-g \mu_B \mathbf{h} \cdot \mathbf{S}$.

Now we are in position to evaluate ${\cal{F}}_{\text{S}}(\mathbf{M})$ as a function of $M$ for two azimuthal angles $\phi = 0$ and $\pi/4$ as a series expansion in $M$ assuming that $a$ is small. The resulting values of $K_C$ are shown in Fig.\,\ref{fig: single_ion}.

To conclude this section we compare in Fig.~\ref{Fig:KC-KU-T} the experimental
data with a theoretical result obtained within the extended Zener model,
discussed above.
In these computations,  $p = 3.3 \times 10^{20}$~cm$^{-3}$, as inferred from the magnitudes of $T_{\mathrm{C}}$ and $M_{\mbox{\tiny S}}$ of Sample A.
Furthermore, we assume  $\sigma = 70$~meV in order to reproduce the experimental magnitude of $K_C$ in the high temperature region.
The divergence between the theoretical and experimental data visible at $T \rightarrow 0$ 
indicates that at $x_{\text{eff}} \simeq 7 \%$ the single-ion anisotropy is too weak to overcome a
large carrier liquid contribution and to explain the sign change of $K_C$ inferred experimentally at low temperatures.
On the other hand, choosing a higher value of $\sigma$ allows to reproduce
this change of sign but the resulting magnitude of $|K_C|$ is much smaller than experimental values.





\section{The interface contribution}
\label{sec:Inerface}

But, perhaps, such a stringent measures are not really required and the need to reproduce the second, the low-$T$ change of sign of $K_C$ is largely apparent.
We note here that the magnitudes of both anisotropy constants, although technically obtained in a correct way, are established upon a very strong assumption of a perfect magnetic uniformity of (Ga,Mn)As, the sole condition under which Eq.~(\ref{Eq:energy}) is valid.
In this section we present a method of the experimental assessment of the previously disregarded contribution in micromagnetic consideration of (Ga,Mn)As brought about by the magnetic phase separation driven by electrostatic inhomogeneities specific to the proximity of metal-insulator transition (MIT).

We start form the notion that there has been a growing number of experimental evidences that this SP-L contribution assumes even a dominant role, determining the magnetic properties of low and very low-$x$ samples - there low magnitude of $p$ is guarantied by low Mn doping \cite{Ye:2017_PRM,Ye:2018_JPCM,Gluba:2017_cond-mat} but more importantly, also in structures with a much higher $x$ where low magnitudes of $p$ results from intentional or unintentional drainage of holes out of the DFS \cite{Sawicki:2010_NP,Chen:2015_PRL,Siusys:2014_NL,Sadowski:2017_Nanoscale}.
In particular, such a situation takes place at the vicinity of the free surface of (Ga,Mn)As layer and near the interface with n-type LT-GaAs buffer.
Formed at these limits two space-charge layers sizably deplete both (Ga,Mn)As edges, forcing the carriers, which in DFS mediate the FM order, to localize.
According to the two-fluid model of electronic states in the vicinity of the Anderson-Mott MIT \cite{Paalanen:1991_PB}, it will be either a weak or a strong localization, depending on the degree of depletion.
This electrical disorder, via enhanced local density fluctuations, sets the ground for a magnetic nanoscale phase separation \cite{Dietl:2008_JPSJ,Richardella:2010_S}.
In such an environment the FM order gets constrained to mesoscopic lengths, being maintained only within these fragments which are visited by (weakly) localized holes. 
These small FM volumes, exert (as an ensemble) SP-L properties and introduce 
magnetic features characteristic to dynamical slow down due to activated processes in the presence of energy barriers and, most importantly for this study, a concave curvature of their Langevine-like magnetic isotherms $m_{SP-L}(H)$ is added to the magnetic response of the remaining metallic part of the sample.

The importance of SP-L admixture depends on the volumetric ratio of the mesoscopic to long-range parts of the sample, so it has to be sizable in very thin layers, in particular when effects in time domain \cite{Chen:2015_PRL} or dependent on the curvature of $M(H)$ are probed.
In a broad view the following aspects have to be taken into consideration.
Firstly, because (Ga,Mn)As is at the vicinity of the MIT, it does not take much to impose the magnetic disorder due to local fluctuation of $p$ in samples with
uniform Mn distribution and flat interfaces \cite{Sawicki:2010_NP,Matsukura:2004_PE,Dietl:2008_JPSJ,Kodzuka:2009_U}.
The volume still richly populated by mobile holes will retain their FM response - in the present case accurately described by Eq.~(\ref{Eq:energy}), whereas the depleted regions will show SP-L response.
Secondly, even crystallographically best and uniformly Mn-doped layers have got two limiting surfaces where hole depletion is likely o occur. Therefore, SP-L effects are expected to surface to a certain degree in every (Ga,Mn)As layer.
Thirdly, the formation of the SP-L disorder is expected not only at low $T$.
Since $T_{\mathrm{C}} \propto x$ in DFS the magnetic phase separation may already start even at moderate temperatures at the edges of large $x$ samples, actually persisting up to a significant fraction of $T_{\mathrm{C}}$ for high quality (optimally annealed) films, or, more generally, up to temperatures comparable to, or even exceeding, $T_{\mathrm{C}}$ in electrically compensated samples \cite{Chen:2015_PRL}.
Lastly, and sadly, the details of the magnetic characteristics of this SP-L component are not exactly known, so they cannot be, even on a phenomenological level, correctly included in the magnetostatic energy considerations. 
Therefore, the analysis of the experimentally established quantities, either as a function of magnetic field, temperature, or time  are no longer expected to yield correct results  describing an "ideal" - magnetically homogenous (Ga,Mn)As, as exemplified recently in the case of Gilbert damping constant \cite{Chen:2015_PRL}.

Below, we present our attempt to assess the low-$T$ magnitude of the "interface-born" SP-L magnetic moment $m_{SP-L}(H)$ in 10~nm thin sample A, and basing on some heuristic arguments we show that indeed a presence of such a contribution may revert the sign of the otherwise negative cubic anisotropy constant if the determination of $K_C$ is based on the time honored approach [Eq.~(\ref{Eq:energy})], originally put forward for an idealistic homogenous "edge-less" (Ga,Mn)As.
This effect is brought about by the same \textsl{concave} curvature of the Langevine-like $m_{SP-L}(H)$ which will either enlarge an already positive $K_C$ or may even change its sign to positive, in so mimicking the low temperature spin reorientation transition $\langle 110 \rangle \leftrightarrow \langle 100 \rangle$ of the biaxial component of the magnetic anisotropy.

\begin{figure*}[th]
	\includegraphics[width=1.9\columnwidth]{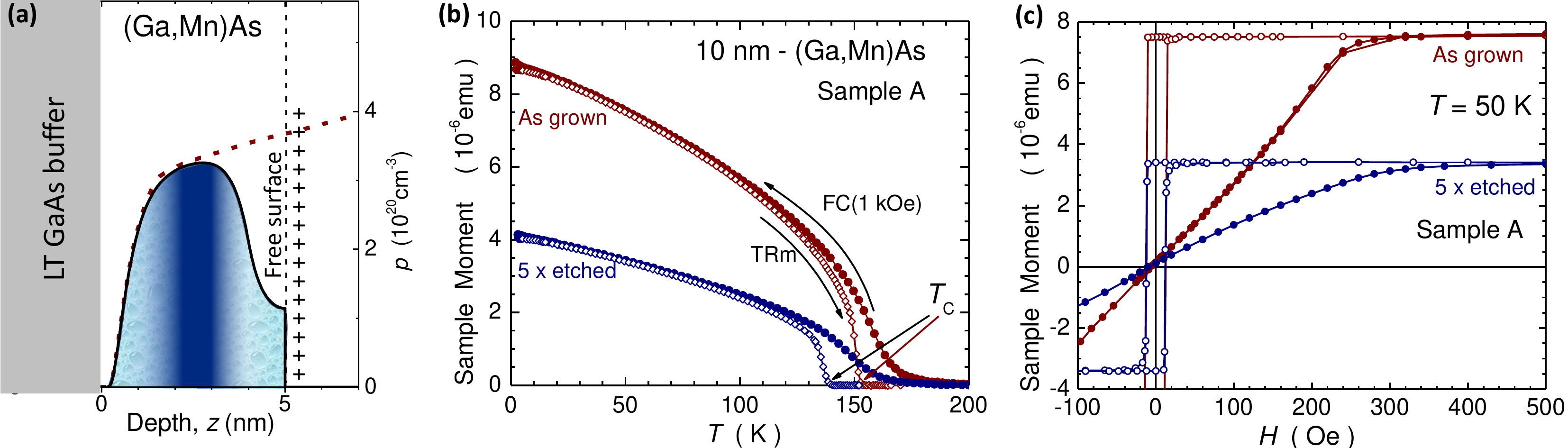}
	\caption{\label{fig:SB_etching}%
(Color online) (a) Thick black line: expected vertical hole density profile $p(z)$ for $d = 5$~nm thin (Ga,Mn)As layer. The dashed line indicates the first 7~nm $p(z)$ in the original (Ga,Mn)As layer before thinning.
An exempt from Fig.~3a in Ref.~\onlinecite{Proselkov:2012_APL}.
The dark shaded part of the profile marks the central part of the layer where uniform ferromagnetic coupling specific to metallic (Ga,Mn)As prevails.
On moving away towards the interfaces the rapidly decreasing $p$ forces the magnetic phase separation, indicated by a light blue texture.
(b) Temperature $T$ and (c) magnetic field $H$ dependent studies of the same samples. The thermoremnant moments (TRm) are acquired for the uniaxial easy orientation (along $[1\bar{1}0]$) on warming after field cooling the samples at $H = 1$~kOe to about 2~K and quenching $H$ to sub-Oe range. Arrows indicate the magnitudes of Curie temperature $T_{\mathrm{C}}$ established upon TRm.
$H$-dependent characteristics for $T=50$~K are shown for both uniaxial easy ($[1\bar{1}0]$ - open symbols) and hard axis ($[110]$ - full symbols) orientations.
}%
\end{figure*}
To evaluate $m_{SP-L}(H)$ in sample A we use the concept of fine thinning of (Ga,Mn)As by multiple etching of the native oxide in HCl \cite{Edmonds:2005_PRB,Olejnik:2008_PRB,Proselkov:2012_APL}.
In particular, we note, following the established upon the same procedure vertical hole density profile $p(z)$ in similar (Ga,Mn)As layer (c.f. Fig.~3a in Ref.\onlinecite{Proselkov:2012_APL}) that by reducing the layer thickness $d$ down to about 5~nm we should be left only with a marginally thin (1-2~nm) high hole density mid-part (slab) of the initial layer, sandwiched between two edge layers (approx. 1.5~nm each) with sizably reduced $p$, as sketched in Fig.~\ref{fig:SB_etching}~a.
Importantly, to assess the role of SP-L component in the initial layer we need only just a so thin layer that (i) it can be regarded as consisting mostly of "two edges" (the top one at the free surface and the bottom one at the interface with the n-type LT GaAs buffer), but that (ii) it remains thick enough to avoid too strong depletion, not present in the original, 10~nm layer.
In our understanding it is the presence of this $\sim 1+$~nm thin central part which guarantees this correspondence.
It needs to be added here that the thinning of the initially studied layer assures us that we deal with the same material as the originally investigated.
Resorting to other, deliberately grown 5 nm layers would greatly reduce the relevance of this exercise as different magnitudes of $x_{\text{eff}}$ and $p$ and/or their different volume distribution are likely.

The effect of etching on magnetic properties is presented in Fig.~\ref{fig:SB_etching}b-c where basic $T$ and $H$ characteristics 
are plotted for the original 10~nm sample A and after 5 consecutive etchings.
After each etching the sample has been left at ambient atmosphere for at least a day to assure a maximum in-diffusion of oxygen and a full oxidation of the topmost $\sim1$~nm of the layer.
The magnitude of the drop of the signal registered in Fig.~\ref{fig:SB_etching} provides also a scaling factor pointing to the intended $\sim5$~nm final thickness after etching.
The $T_{\mathrm{C}}$ of the thinned layer dropped in comparison to the original sample by about 13~K, a number consistent with the reported in Ref.~\onlinecite{Proselkov:2012_APL} reduction at the same range of thicknesses, substantiating the use of the finding established there. 
Another worth mentioning feature of the magnetic data in Fig.~\ref{fig:SB_etching} is the excellent uniaxial behavior exhibited also after thinning. 
This shows that after the full five cycles of etching and oxidations the layer preserved its pristine 
micromagnetic properties.

On the other hand one can evidently note a change of the curvature of $ m_{[110]}(H)$ which took place upon thinning.
We assign this change to a sizable reduction of this fraction of the initial layer which was occupied by mobile holes (and exhibited uniform magnetization characterized by $K_C<0$) in expense of exhibiting concave curvature Langevin-like $m(H)$ originating from the depleted interfaces.
According to $p(z)$ profile presented in Fig.~\ref{fig:SB_etching}a these depleted volumes take up to about $f = 2/3$ of the whole volume of the thinned layer.
Now, using low-$T$ magnetic data, we will evaluate $m_{SP-L}(H)$ in the original sample A at 5~K.

\begin{figure}
	\includegraphics[width=0.99\columnwidth]{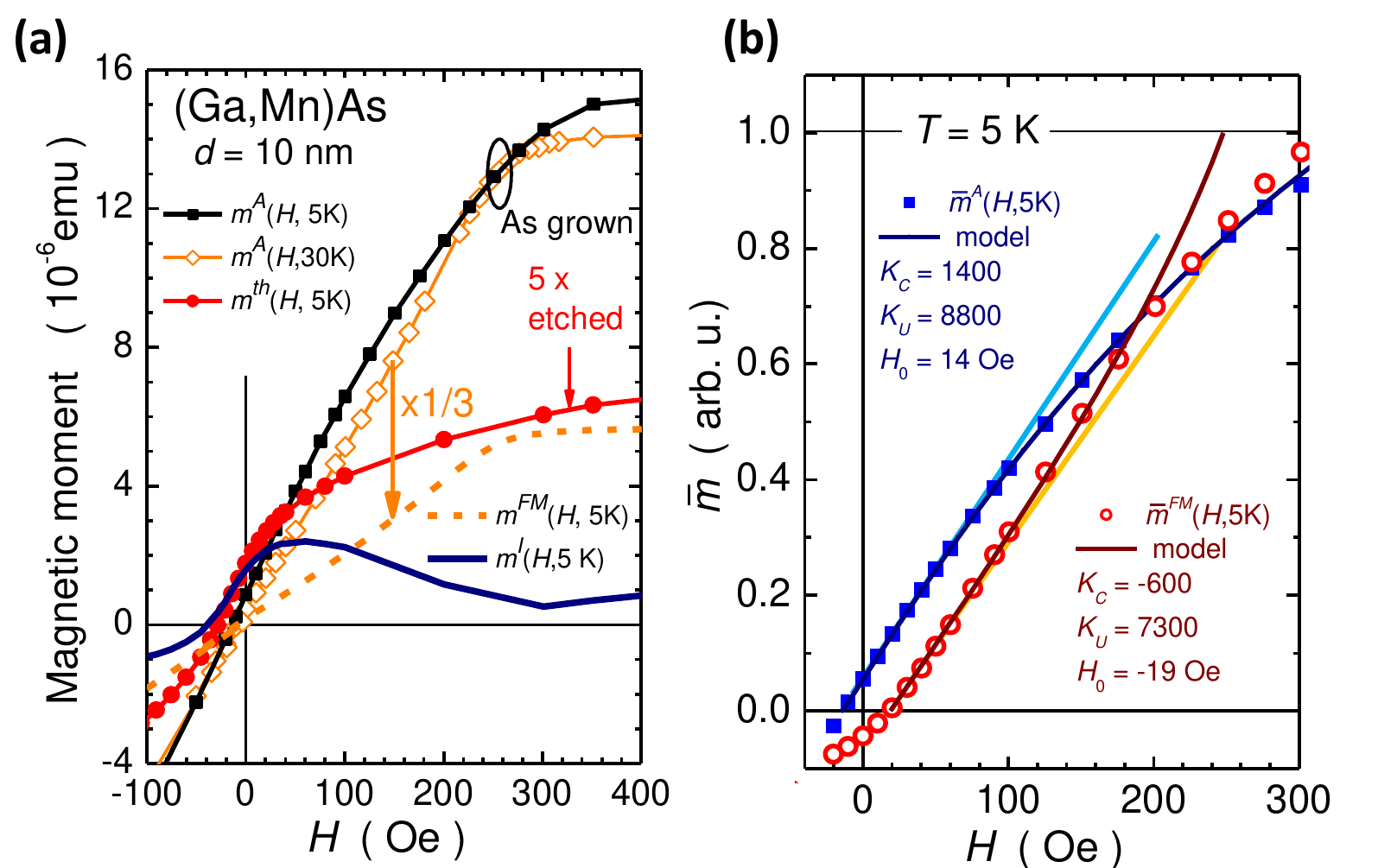}
	\caption{\label{fig:SDGZ} (Color online) 
(a) Illustration of the method used to establish low temperature magnetic field $H$ response specific to interfaces  in 10~nm thin sample A (the thick navy line). All measurements are performed along $[110]$, the uniaxial hard direction. The labeling of the data sets corresponds to the notation used in the text.
(b) The same sample. A comparison between the original hard axis magnetization curve (solid squares) and the specific to ferromagnetic (uniform) part of the sample (open circles).
The background thick lines of lighter shades indicate the initial slope of each $\bar{m}(H)$  and serve as references to ease the identification of the opposite mid-field curvatures. The solid lines of matching colors represent the modeling of the data by Eq.~(\ref{Eq:Hm110}). 
The magnitudes of the established uniaxial $K_U$ and cubic $K_C$ anisotropy constants are denoted in the panel (in units of erg/cm$^{-3}$), together with values of a required extra horizontal shift $H_0$ to align modeled curves with the experimental points at $m(H) \simeq 0$.}%
\end{figure}

Corresponding uniaxial hard axis magnetization curves measured at 5~K for sample A $ m_{[110]}^A (H,5\mathrm{~K})$ and after thinning $ m_{[110]}^{th} (H,5K)$ are presented in Fig.~\ref{fig:SDGZ}a, indicating, similarly to 50~K, a more concave character of $ m_{[110]}(H)$ of the thinned layer (bullets).
On the other hand, the added to this figure $ m_{[110]}(H)$ of the same sample measured at 30~K (open diamond), in an accordance with the data presented in Figs.~\ref{Fig:ExampleM-H} and~\ref{Fig:KC-KU-T}, exhibits a convex curvature.
This points to rather weak SP-L contribution at this $T$ and allow us to take this $m(H)$ as the reference data which represent the low-$T$ $ m_{[110]}(H)$ of an ideal edge-less (Ga,Mn)As with $K_C < 0$.
We substantiate this choice upon the results presented in Fig.~\ref{Fig:KC-KU-T}, where established at higher $T$ for this sample values of $K_C$ continue to grow more negative on lowering $T$ until just about 30~K.


However, in the etched layer this response should be exerted by about a third $(1-f)$ of its volume, so by scaling it down 3 times (the dashed line in Fig.~\ref{fig:SDGZ}a) and subtracting from $m_{[110]}^{th}(H,5K)$ we obtain the required experimental estimation of the interface-born contribution to $ m(H)$ at 5~K, $m_{[110]}^I (H,5K)$, marked as the thick solid line in Fig.~\ref{fig:SDGZ}a.
The magnitude of $m_{[110]}^{I}(H,5K)$ is rather weak, particularly when compared to the saturation values of $ m_{[110]}^A (H,5\mathrm{~K})$, so it could be regarded as a secondary contribution to any leading characteristics of the material.
However, and in an accordance to our expectations, this $m_{[110]}^{I}(H,5K)$ exhibits a very strong concave curvature at weak magnetic fields, and so a presence of such a contribution sizably impair the determination of $K_C$.

Having evaluated the SP-L contribution we are in position to calculate the FM response of sample A at 5~K, \textsl{i.e.} as it would be if the sample hadn't had depleted regions near its interfaces $ m_{[110]}^{FM} (H,5\mathrm{~K}) = m_{[110]}^A (H,5\mathrm{~K}) - m_{[110]}^{I}(H,5K)$.
The original and calculated $m(H)$ are plotted in Fig.~\ref{fig:SDGZ}b in relative units $\bar{m}$, marked by full and open symbols, respectively.
Thick solid lines of lighter shades represent their initial slopes (established around $\bar{m}\simeq 0$) and guide the eyes to indicate the opposite curvatures of both dependencies in the mid-field region.
Importantly, the new $\bar{m}(H)$, the $ m_{[110]}^{FM} (H,5\mathrm{~K})$, exhibits now a clear \textsl{upward} shift in the mid-field range, which according to the general model of magnetic anisotropy in (Ga,Mn)As [Eq.~(\ref{Eq:Hm110})] corresponds to a negative $K_C$.
We can now evaluate its magnitude using the same method as employed before in section~\ref{sec:ExpDet}.
The established this way magnitude of $K_C (5$~K)$ = -600$~erg/cm$^{-3}$, represented in Fig.~\ref{Fig:KC-KU-T} by a star,  is most likely still far from being a precise one, but is undoubtedly opposite to that one obtained form the analysis of bare $ \bar{m}_{[110]}^A (H,5\mathrm{~K})$, and it corresponds much better to the expectations brought about by the elaborated in the previous section extended Zener model of FM in (Ga,Mn)As.
It needs to be added that the final outcome of our procedure, in particular  the negative sign of $K_C$, does not depend on the exact choice of $(1-f)$ from 0.25 to 0.5.

We finally comment on the strangely negative values of  $ \bar{m}_{[110]}^{FM} (H,5\mathrm{~K})$ at $H \simeq 0$ (Fig.~\ref{fig:SDGZ}b).
In our view this is a result of yet another characteristic feature of the inhomogeneous constitution  of (Ga,Mn)As.
We do not elaborate on this issue here, it a subject of an in-depth independent study.
We remark only that when (Ga,Mn)As is measured within the parameter space corresponding to the formation of SP-L phase, this fragment of $M(H)$ where the magnetization reversal takes place is strongly dependent on the rate at which the magnetic field is swept, particularly at low $T$ \cite{ Hrabovsky:2002_APL}, what is a characteristic feature of dynamical slow down due to activated processes in the presence of energy barriers.
Since thinning and/or electrical compensation in (Ga,Mn)As promotes a growth of $f$ towards unity, a low-$T$ increase of experimentally established coercivity $H_C$ is indeed expected in thinned sample.
This enlargement can be noticed in Fig.~\ref{fig:SDGZ}a for $ m_{[110]}^{th} (H,5\mathrm{~K})$ and it is this enlarged coercivity with respect to more homogeneous sample A at 5 and 30~K that is the source of the slight down-shift of  $ m_{[110]}^{FM} (H,5\mathrm{~K})$.
Actually, the rapidly growing magnitude of $H_C$ upon further thinning (not shown) is the source of the second (practical) reason why the evaluation of SP-L contribution stopped at 5~nm.
Nevertheless, even in such a case the magnitudes of anisotropy constants can be still evaluated within the frame of the method described in section~\ref{sec:ExpDet} by introducing an artificial extra parameter allowing to align modeled by Eq.~(\ref{Eq:Hm110}) $\bar{m}_{[110]}(H)$ with the experimental points at $\bar{m}(H) =0$.

We summarize this section by noting why the $\langle 110\rangle$ cubic easy axes in (Ga,Mn)As might had gone unnoticed before.
Surely, as with all the physical properties of DFS the samples must have the right magnitudes of $p$, $T$, and $x_{\text{eff}}$ to grant the adequate balance of the relevant terms describing the free energy of the system.
Secondly, as shown just above, the sample must be of a high magnetic uniformity to suppress the detrimental for "negative $K_C$" contribution from SP-L phase, at least in not-too thin layers and/or having strongly reduced interface depletion.
But it is highly unlikely that our samples are the first which meet the pointed above criteria.
So, we want to turn the attention to a far more down-to-earth reason: the details of experimental procedure.
As presented in the study, the $\langle 110\rangle$ cubic easy axes have been observed exclusively for not-fully-annealed samples (\textsl{i.e.} in the middle of a small step annealing process) and at temperatures above 15~K and well below $T_{\mathrm{C}}$ - that is outside the envelope of typical conditions at which (Ga,Mn)As is tested or investigated.
For example, a typical assessment of $M_{\mbox{\tiny S}}$ is made at $T \leq 5$~K.
Here SP-L contribution really gets strength.
On the other hand, the $T_{\mathrm{C}}$ is frequently established from thermoremnant measurement which does not hint on the exact orientation of the cubic easy axes particularly at $T \rightarrow Tc$ where $|K_C| < |K_U|$.
So, it is very likely that the existence of the negative $K_{C}$ in (Ga,Mn)As might have been simply overlooked due to a too routine approach to the material characterization.

\section{Conclusions}

Magnetic anisotropy of of carefully prepared high quality thin layers of (Ga,Mn)As have been studied as a function of temperature and hole concentration both experimentally and theoretically.
On the account of magnetic and ferromagnetic resonance studies it has been convincingly evidenced that within a certain range of $p$ and $T$ parameter space the cubic component to the in-plane magnetic anisotropy assumes the $\langle110\rangle$ easy exes.
Accordingly, outside this frame, the cubic anisotropy reverts to ubiquitously reported $\langle100\rangle$ directions, indicating an oscillating dependence of anisotropy constant on $p$ and $T$.
These for the very first time observed features qualitatively confirm the relevant predictions of the $p$--$d$ Zener model of ferromagnetism in dilute ferromagnetic semiconductors \cite{Dietl:2000_S}, what can be taken as strong experimental support for the model.
In particular, even a quantitative agreement has been obtained in high-$T$ and high-$p$ part of the data when the developed here more advanced form of the Zener model, which takes both the single-ion magnetic anisotropy of Mn species and the disorder into account, is applied.
However, even in this advanced form the model cannot quantitatively reproduce the low-$p$ and low-$T$ rotation back of the easy axis to $\langle100\rangle$ orientations.
We note, however,  that in this regime, the magnitude and sign of apparent $K_C$ values might be fraught with an error brought about by  the magnetic phase separation into ferromagnetic and superparamagnetic-like regions.
The phase separation is, in turn, driven by  electrostatic disorder specific to the proximity of the metal-insulator transition, particularly in interfacial regions of the layers, in which hole liquid is depleted and, thus, prone to localization.
The key point of our reasoning is that the Langevine-like superparamagnetic response, by introducing the same concave magnetization curvature  as is expected for the $\langle100\rangle$ oriented cubic term, forces any data analyzing procedure to yield more positive values for $K_C$, as it would in the absence of this contribution.
To substantiate our claims we have presented the experimental procedure that allows to assess the magnitude of this detrimental paramagnetic contribution to $m(H)$, and after correcting the data we show that indeed the negative sign of $K_C$ is obtained.

Our results substantiate, therefore, the importance of the interfaces in the understanding of physical processes that take place in thin layers of DFS, and (Ga,Mn)As in particular.
Despite the fact that this has been known for some time now \cite{Nishitani:2010_PRB,Sawicki:2010_NP,Chen:2015_PRL}, this study shows how the effects induced by interfacial properties of (Ga,Mn)As preclude detection of very relevant and important for the community findings.

It may well turn to be a dominant, also as a volume-born contribution, in electrically compensated and characterized by low magnitude of $T_{\mathrm{C}} / x$ ratio samples \cite{Chen:2015_PRL},  as it is already established to be the dominating magnetic response in very low Mn content (III,Mn)V samples \cite{Siusys:2014_NL,Chen:2015_PRL,Sadowski:2017_Nanoscale,Ye:2017_PRM,Ye:2018_JPCM,Gluba:2017_cond-mat}.
Interestingly, whereas the electrical disorder, particularly in electrically compensated samples, was originally expected  to smooth out the oscillatory behavior of the cubic anisotropy \cite{Dietl:2001_PRB}, the present findings point out to a new mechanism by which the electrical inhomogeneity affects the magnetic constitution of DFS.

\section*{Acknowledgments}

This study has been supported by the National Science Centre (Poland) through PRELUDIUM Grant No. 2012/05/N/ST3/03147 and MAESTRO Grant No. 2011/02/A/ST3/00125. EC Network SemiSpinNet (PITN-GA-2008-215368) contribution is also acknowledged.
The International Centre for Interfacing Magnetism and Superconductivity with
Topological Matter project is carried out within the International Research Agendas
programme of the Foundation for Polish Science co-financed by the European Union
under the European Regional Development Fund.

\bibliography{ref_30Jan2018}

\end{document}